\documentclass[12pt]{article}
\usepackage{amsfonts,amssymb,amsmath,mathrsfs}
\usepackage{hyperref}
\newcommand{\beq}{\begin{equation}}
\newcommand{\eeq}{\end{equation}}
\newcommand{\vp}{\vphantom}

\newcommand{\wt}{\widetilde}
\newcommand{\wh}{\widehat}
\newcommand{\al}{\alpha}
\newcommand{\bt}{\beta}
\newcommand{\g}{\gamma}
\newcommand{\de}{\delta}

\newcommand{\mc}{\mathcal}
\newcommand{\mr}{\mathrm}

\begin{document}
\begin{center}
\textbf{\Large Spinor description of\\[0.3cm] $D=5$ massless low-spin gauge fields}\\[0.3cm]
{\large D.V.~Uvarov\footnote{E-mail: d\_uvarov@\,hotmail.com}}\\[0.2cm]
\textit{NSC Kharkov Institute of Physics and Technology,}\\ \textit{61108 Kharkov, Ukraine}\\[0.5cm]
\end{center}

\begin{abstract}
Spinor description for the curvatures of $D=5$ Yang-Mills,
Rarita-Schwinger and gravitational fields is elaborated.
Restrictions imposed on the curvature spinors by the dynamical
equations and Bianchi identities are analyzed. In the absence of
sources symmetric curvature spinors with $2s$ indices obey
first-order equations that in the linearized limit reduce to
Dirac-type equations for massless free fields. These equations
allow for a higher-spin generalization similarly to $4d$ case.
Their solution in the form of the integral over Lorentz-harmonic
variables parametrizing coset manifold $SO(1,4)/(SO(1,1)\times
ISO(3))$ isomorphic to the three-sphere is considered.
Superparticle model that contains such Lorentz harmonics as
dynamical variables, as well as harmonics parametrizing the
two-sphere $SU(2)/U(1)$ is proposed. The states in its spectrum
are given by the functions on $S^3$ that upon integrating over the
Lorentz harmonics reproduce on-shell symmetric curvature spinors
for various massless supermultiplets of $D=5$
space-time supersymmetry.
\end{abstract}

\section{Introduction}

Spinor approach to the description of $4d$ gravitational field
initiated by R.~Penrose \cite{AP'60} can be extended in the
uniform way to other gauge fields that manifests itself in the
construction of the contour integral solutions of free massless
field equations \cite{IJTP'68} using the features of two-component
$SL(2,\mathbb C)$ spinors and $SU(2,2)$ twistors
\cite{Penrose'67}. Symmetric curvature spinors with $2s$
indices that appear in such an approach generalize linearized Weyl curvature spinor ($s=2$) and were
shown later to play an important role in the formulation of higher spin gauge theories extending Einstein
\cite{Vasiliev-PLB} and Weyl \cite{Linetsky} gravity theories. 
Generalizations of such a construction relying on twistors related
to higher-dimensional conformal symmetries encounter difficulties
(see, e.g., \cite{SWolf}, \cite{MRETCh}) because of the more
complicated algebra of multicomponent spinors and the fact that in
dimensions greater than four the requirement of conformal
invariance turns out very restrictive.

That is why an approach, in which only Lorentz invariance is manifest, looks more preferable. In Ref.~\cite{DelducGS} there was constructed on-shell integral representation\footnote{Analogous integral representation for solutions of the equations for $4d$ massless arbitrary spin fields had been independently proposed in \cite{Bandos'90}.}
for the curvatures of massless free fields in dimensions $D=3,4,6,10$ that uses Lorentz vector and spinor harmonics\footnote{Recall that Lorentz harmonics were introduced in \cite{Sokatchev'86} as a generalization of the harmonic variables for compact groups \cite{GIKOS}. Spinor harmonics were introduced in \cite{DelducGS}, \cite{GalperinHS}, \cite{BZ-JETP'91} (For early attempts on introducing spinor harmonics see, e.g., \cite{NPS-NPB296},\cite{Solomon}). Let us note that in $D=4$ spinor harmonic matrix can be identified with the normalized dyad \cite{Newman}. The original motivation for considering such objects was connected with the hope to solve the problem of covariant quantization of the Green-Schwarz superstrings using the spinor harmonics. Recently spinor variables that could be related to gauge-fixed Lorentz harmonics were introduced in \cite{Cheung'09}, \cite{Boels}, \cite{Caron-Huot} and used to elaborate on the spinor-helicity formalism for higher-dimensional gauge fields and spinor representation of the scattering amplitudes.}
so that the Lorentz-symmetry is built in. The integrand is a function of the rectangular block of the spinor harmonics matrix and the projection of the space-time coordinate vector onto a null vector constructed out of those spinor harmonics. The form of the harmonic measure ensures invariance of the integral under $SO(1,1)\times ISO(D-2)$ gauge symmetry so that the integration is actually performed over the sphere $S^{D-2}$. Resultant curvature spinors are functions of the space-time coordinates and satisfy Dirac-type equations.

In the present paper we generalize this approach to the $5d$ case.
To this end we work out in detail the spinor description for $D=5$
Yang-Mills (YM), Rarita-Schwinger (RS) and gravitational fields.
Their irreducible curvature spinors are characterized and contents
of the dynamical equations and Bianchi identities written in terms
of the curvature spinors is analyzed. It is shown that in the
absence of sources there remain non-zero only totally symmetric
curvature spinors with $2s$ indices that satisfy first-order differential equations. In the
linearized limit they are shown to reduce to the equations for the corresponding
linearized curvatures of free massless fields and allow
straight-forward higher-spin generalization. These equations may
also be viewed as a $5d$ generalization of $4d$ first-order
equations \cite{McCallum}, \cite{PR}, \cite{Vasiliev-PLB+AP} obeyed by the generalized
Weyl curvature spinors. It
should be noted that in the gravitiational literature the spinor form of the $D=5$ Weyl tensor was
considered in \cite{DeSmet}\footnote{See also \cite{Godazgar},
where additionally spinor form of the Maxwell curvature was
discussed.} and the decomposition of the Riemann tensor on the
irreducible curvature spinors was obtained in \cite{Garcia}. Independently in Ref.~\cite{SS} spinor form of the linearized Weyl tensor and its higher-spin counterparts was used in the unfolded formulation of equations of motions for free fields on $AdS_5$ that correspond to the unitary irreducible representation of certain higher-spin generalization of $su(2,2)$ algebra and its supersymmetric extension pertinent to $AdS_5/CFT_4$ duality.\footnote{Non-linear equations for symmetric massless higher-spin fields on $D$-dimensional $(A)dS$ space were constructed in \cite{Vasiliev2003} using vectorial generating elements of the underlying higher-spin algebra. We note, however, the importance of spinorial realization of higher-spin algerbas and field dynamics in view of possible supersymmetric generalizations (for $D=5$ case see \cite{Vasiliev-Alkalaev}).}

Then we present integral representation for Weyl curvature spinors of arbitrary spin free fields that makes use of $D=5$ Lorentz harmonics\footnote{Lorentz harmonics in dimension $D=5$ were first introduced in \cite{Sokatchev'89}, although there another coset realization was considered.} and solves Dirac-type equations analogously to the cases considered in Ref.~\cite{DelducGS}.

In the last part of the paper we propose a superparticle model characterized by the set of simple irreducible constraints, whose first-quantized states are given by the multiplets of harmonic functions that correspond to Weyl curvature spinors of the fields from various $D=5$ supermultiplets. The model includes $D=5$ spinor harmonics parametrizing the coset $SO(1,4)/(SO(1,1)\times ISO(3))$ and harmonics of Ref.~\cite{GIKOS} parametrizing the two-sphere among the dynamical variables. In the appendices are collected relevant properties of spinors, $\g-$matrices and Lorentz harmonics.

\section{Curvature spinors of $D=5$ massless gauge fields and equations for them}

In this section we find $Spin(1,4)$ spinor form of the curvatures for YM, RS and gravitational fields, study in detail the restrictions imposed on the curvature spinors by field equations and Bianchi identities and consider their linearized limit.

\subsection{Spinor form of YM field equations and Bianchi identities}

We start with the YM case.\footnote{For simplicity we suppress the
gauge algebra indices.}  YM curvature tensor $F_{mn}$ is
converted into the spinor form by contracting vector indices with
those of $D=5$ $\g-$matrices
\beq\label{ym-curvature-spinor}
F_{\al[2]\bt[2]}=C_{\al_1\bt_1}F_{\al_2\bt_2}-C_{\al_1\bt_2}F_{\al_2\bt_1}-C_{\al_2\bt_1}F_{\al_1\bt_2}+C_{\al_2\bt_2}F_{\al_1\bt_1}.
\footnote{For the spinor indices we
adopt the shorthand notation that a number in square brackets
following an index stands for the group of indices equal to that
number that are antisymmetrized with unit weight. Similarly a number in round brackets following an index
denotes the group of indices symmetrized with unit weight.}
\eeq
Charge conjugation matrix $C_{\al\bt}$ that appears on the r.h.s.
is antisymmetric (see Appendix A for spinor algebra in 5
dimensions). For real Yang-Mills field symmetric curvature spinor
$F_{\al\bt}=F_{\bt\al}$ satisfies Hermiticity condition
\beq
(F_{\al\bt})^\dagger=\g^{0\bt\g}F_{\g\de}\g^{0\de\al}
\eeq
and
thus has 10 real components. Spinor and vector
representations of the curvature can be converted into one another using the relations
\beq
F_{\al(2)}=-\frac12\g^{mn}\vp{\g}_{\al(2)}F_{mn},\quad
F_{mn}=-\frac14\g_{mn}\vp{\g}^{\al(2)}F_{\al(2)}.
\eeq
Vacuum
Yang-Mills equations
\beq\label{vYMeq} \nabla^{m}F_{mn}=0
\eeq
in
spinor representation read
\beq\label{YMveq}
\nabla_{\al}\vp{\nabla}^{\g}F_{\g\bt}-\nabla_{\bt}\vp{\nabla}^{\g}F_{\g\al}=0,
\eeq
where
$\nabla_{\al}\vp{\nabla}^{\bt}=\g^{m}\vp{\g}_{\al}\vp{\g}^{\bt}\nabla_{m}$
is the spinor form of the YM covariant derivative. Bianchi
identity $\nabla\wedge F=0$ transforms into equation
\beq\label{YMBianchi}
\nabla_{\al}\vp{\nabla}^{\g}F_{\g\bt}+\nabla_{\bt}\vp{\nabla}^{\g}F_{\g\al}=0.
\eeq
Combining Eqs.~(\ref{YMveq}) and (\ref{YMBianchi}) yields
that the curvature spinor satisfies
\beq
\nabla_{\al}\vp{\nabla}^{\g}F_{\g\bt}=0.
\eeq
In the linearized
limit one obtains free spin-1 field equation
\beq\label{spin1eq}
\partial_{\al}\vp{\partial}^{\g}F_{\g\bt}=0.
\eeq

\subsection{Spinor form of RS field equations and Bianchi identities}

In general the RS field $\Psi_{m\g}\vp{\Psi}^a$ carries Lorentz vector and spinor indices, we additionally endow it with the index $a$ to account for the case of gravitini fields in $N-$extended supergravity multiplets. Curvature spin-tensor
\beq\label{psicurvature-vector}
\Phi_{mn\g}\vp{\Phi}^a=\partial_{m}\Psi_{n\g}\vp{\Psi}^a-\partial_{n}\Psi_{m\g}\vp{\Psi}^a
\eeq
similarly to the YM curvature tensor can be converted to the spinor form
\beq
\Phi_{\al[2]\bt[2]\g}\vp{\Phi}^a
=C_{\alpha_1\beta_1}\Phi_{\alpha_2\beta_2|\g}\vp{\Phi}^a-C_{\alpha_1\beta_2}\Phi_{\alpha_2\beta_1|\g}\vp{\Phi}^a-C_{\alpha_2\beta_1}\Phi_{\alpha_1\beta_2|\g}\vp{\Phi}^a+C_{\alpha_2\beta_2}\Phi_{\alpha_1\beta_1|\g}\vp{\Phi}^a.
\eeq
Curvature spinor
\beq
\Phi_{\al(2)|\bt}\vp{\Phi}^a=\frac12(\partial_{\al_1}\vp{\partial}^{\de}\Psi_{\al_2\de|\bt}\vp{\Psi}^a+\partial_{\al_2}\vp{\partial}^{\de}\Psi_{\al_1\de|\bt}\vp{\Psi}^a),\quad\Psi_{\al[2]|\bt}\vp{\Psi}^a=\g^{m}\vp{\g}_{\al_1\al_2}\Psi_{m\bt}\vp{\Psi}^a
\eeq
is symmetric in the first two indices that are therefore separated from the last one by vertical line and satisfies Hermiticity condition
\beq
(\Phi_{\al\bt|\g}\vp{\Phi}^a)^\dagger=\Omega_{ab}\g^{0\al\lambda}\g^{0\bt\mu}\Phi_{\lambda\mu|\rho}\vp{\Phi}^b\g^{0\rho\g},
\eeq
where $\Omega_{ab}=-\Omega_{ba}$ is the symplectic metric tensor.\footnote{Recall that allowed $R-$symmetry groups in $D=5$ supergravities are $USp(N)$ with $N=2,4,6,8$.} It has $40N$ components and can be presented as the sum of two summands having $20N$ components each
\beq\label{Phi-decomp}
\Phi_{\al(2)|\bt}\vp{\Phi}^a=\Phi_{\al_1\al_2\bt}\vp{\Phi}^a+\widehat\Phi_{\al(2)|\bt}\vp{\Phi}^a.
\eeq
The first term is totally symmetric in the spinor indices
\beq\label{Phi-symmetric}
\Phi_{\al(3)}\vp{\Phi}^a=\frac13\left(\Phi_{\al_1\al_2|\al_3}\vp{\Phi}^a+\Phi_{\al_2\al_3|\al_1}\vp{\Phi}^a+\Phi_{\al_3\al_1|\al_2}\vp{\Phi}^a\right),
\eeq
while the second
\beq
\widehat\Phi_{\al(2)|\bt}\vp{\Phi}^a=\frac23\left(\Phi_{\al_1\al_2|\bt}\vp{\Phi}^a-\frac12\Phi_{\al_2\bt|\al_1}\vp{\Phi}^a-\frac12\Phi_{\bt\al_1|\al_2}\vp{\Phi}^a\right)
\eeq
has the same symmetry as $\Phi_{\al(2)|\bt}\vp{\Phi}^a$ but its totally symmetrized part vanishes
\beq\label{hatphisym}
\widehat\Phi_{\al_1\al_2|\bt}\vp{\Phi}^a+\widehat\Phi_{\al_2\bt|\al_1}\vp{\Phi}^a+\widehat\Phi_{\bt\al_1|\al_2}\vp{\Phi}^a=0.
\eeq
RS equation
\beq
\g^{klm}\vp{\g}_{\al}\vp{\g}^{\bt}\Phi_{lm\bt}\vp{\Phi}^a=0
\eeq
transforms into the following equation for $\widehat\Phi_{\al(2)|\bt}\vp{\Phi}^a$:
\beq
\widehat\Phi_{\al_1\al_2|\bt}\vp{\Phi}^a-\widehat\Phi_{\al_1\bt|\al_2}\vp{\Phi}^a=0.
\eeq
Combined with (\ref{hatphisym}) it amounts to
\beq\label{RSeqn'}
\widehat\Phi_{\al(2)|\bt}\vp{\Phi}^a=0.
\eeq
Bianchi identity for the curvature spin-tensor (\ref{psicurvature-vector}) in the spinor form reads
\beq
\varrho_{\al(2)|\bt}\vp{\varrho}^a=\partial_{\al_1}\vp{\partial}^{\de}\Phi_{\de\al_2|\bt}\vp{\Phi}^a+\partial_{\al_2}\vp{\partial}^{\de}\Phi_{\de\al_1|\bt}\vp{\Phi}^a=0.
\eeq
Since its symmetry is the same as that of $\Phi_{\al(2)|\bt}\vp{\Phi}^a$, decomposition analogous to (\ref{Phi-decomp}) applies also to $\varrho_{\al(2)|\bt}\vp{varrho}^a$. Provided RS equation (\ref{RSeqn'}) is satisfied, the totally symmetric part $\varrho_{\al(3)}\vp{varrho}^a$ is expressed in terms of the totally symmetric part of the curvature spinor (\ref{Phi-symmetric})
\beq
\varrho_{\al(3)}\vp{varrho}^a=\frac23\left(\partial_{\al_1}\vp{\partial}^{\de}\Phi_{\de\al_2\al_3}\vp{\Phi}^a+\partial_{\al_2}\vp{\partial}^{\de}\Phi_{\de\al_3\al_1}\vp{\Phi}^a+\partial_{\al_3}\vp{\partial}^{\de}\Phi_{\de\al_1\al_2}\vp{\Phi}^a\right)=0.
\eeq
For another part $\widehat\varrho_{\al(2)|\bt}\vp{varrho}^a$, whose total symmetrization identically vanishes, one obtains
\beq
\widehat\varrho_{\al(2)|\bt}\vp{varrho}^a=-\frac23\left(\partial_{\bt}\vp{\partial}^{\de}\Phi_{\de\al_1\al_2}\vp{\Phi}^a-\frac12\partial_{\al_1}\vp{\partial}^{\de}\Phi_{\de\al_2\bt}\vp{\Phi}^a-\frac12\partial_{\al_2}\vp{\partial}^{\de}\Phi_{\de\al_1\bt}\vp{\Phi}^a\right)=0.
\eeq
If further take into account the following representation for the derivative of the totally symmetric part of the curvature (\ref{Phi-symmetric})
\beq
\partial_{\al}\vp{\partial}^{\de}\Phi_{\de\bt_1\bt_2}\vp{\Phi}^a=\frac12\varrho_{\al\bt_1\bt_2}\vp{varrho}^a-\widehat\varrho_{\bt_1\bt_2|\al}\vp{varrho}^a,
\eeq
we conclude that
\beq\label{spin32eqn}
\partial_{\al}\vp{\partial}^{\de}\Phi_{\de\bt(2)}\vp{\Phi}^a=0,
\eeq whenever RS equation and Bianchi identities are fulfilled.

\subsection{Spinor form of gravitational field equations and Bianchi identities}

This section culminates in a discussion of $D=5$ Einstein gravity. Spinor form of the Riemann tensor
\beq
\begin{array}{l}
R_{\al[2]\bt[2]\g[2]\de[2]}=\\[0.2cm]
+C_{\al_1\bt_1}\left(C_{\g_1\de_1}R_{\al_2\bt_2|\g_2\de_2}-C_{\g_1\de_2}R_{\al_2\bt_2|\g_2\de_1}-C_{\g_2\de_1}R_{\al_2\bt_2|\g_1\de_2}+C_{\g_2\de_2}R_{\al_2\bt_2|\g_1\de_1}\right)\\[0.2cm]
-C_{\al_1\bt_2}\left(C_{\g_1\de_1}R_{\al_2\bt_1|\g_2\de_2}-C_{\g_1\de_2}R_{\al_2\bt_1|\g_2\de_1}-C_{\g_2\de_1}R_{\al_2\bt_1|\g_1\de_2}+C_{\g_2\de_2}R_{\al_2\bt_1|\g_1\de_1}\right)\\[0.2cm]
-C_{\al_2\bt_1}\left(C_{\g_1\de_1}R_{\al_1\bt_2|\g_2\de_2}-C_{\g_1\de_2}R_{\al_1\bt_2|\g_2\de_1}-C_{\g_2\de_1}R_{\al_1\bt_2|\g_1\de_2}+C_{\g_2\de_2}R_{\al_1\bt_2|\g_1\de_1}\right)\\[0.2cm]
+C_{\al_2\bt_2}\left(C_{\g_1\de_1}R_{\al_1\bt_1|\g_2\de_2}-C_{\g_1\de_2}R_{\al_1\bt_1|\g_2\de_1}-C_{\g_2\de_1}R_{\al_1\bt_1|\g_1\de_2}+C_{\g_2\de_2}R_{\al_1\bt_1|\g_1\de_1}\right)
\end{array}
\eeq
can be viewed as the 'square' of the corresponding spinor form of the YM curvature (\ref{ym-curvature-spinor}). Riemann curvature spinor $R_{\al(2)|\bt(2)}$ is symmetric in the first and the second pairs of indices and under their interchange $R_{\al(2)|\bt(2)}=R_{\bt(2)|\al(2)}$. Inverse relation gives the Riemann tensor in terms of the Riemann spinor
\beq\label{rim-vec-spin}
R_{klmn}=\frac{1}{16}\g_{kl}\vp{\g}^{\al(2)}\g_{mn}\vp{\g}^{\bt(2)}R_{\al(2)|\bt(2)}.
\eeq
Taking trace in the two vector indices in (\ref{rim-vec-spin}) expresses Ricci tensor via the Riemann curvature spinor
\beq
R_{km}=R_{kpm}\vp{R}^{p}=-\frac18\g_{k}\vp{\g}^{\al_1\bt_1}\g_{m}\vp{\g}^{\al_2\bt_2}R_{\al_1\al_2|\bt_1\bt_2}+\frac18\eta_{km}R_{\al\bt|}\vp{R}^{\al\bt}.
\eeq
Thus the spinor form of the Ricci tensor is
\beq\label{ricci-spinor-form}
\begin{array}{rl}
R_{\al[2]\bt[2]}=&R_{\al_1\bt_2|\al_2\bt_1}-R_{\al_1\bt_1|\al_2\bt_2}\\[0.2cm]
-&\frac12\left(C_{\al_1\bt_1}R_{\al_2\g|}\vp{R}^{\g}\vp{R}_{\bt_2}-C_{\al_1\bt_2}R_{\al_2\g|}\vp{R}^{\g}\vp{R}_{\bt_1}-C_{\al_2\bt_1}R_{\al_1\g|}\vp{R}^{\g}\vp{R}_{\bt_2}+C_{\al_2\bt_2}R_{\al_1\g|}\vp{R}^{\g}\vp{R}_{\bt_1}\right).
\end{array}
\eeq
Further tracing gives scalar curvature
\beq
R=\frac12R_{\al\bt|}\vp{R}^{\al\bt}.
\eeq

Consider the properties of the Riemann curvature spinor. $R_{\al(2)|\bt(2)}$ has 55 components, as the Riemann tensor, and is reducible. It can be represented as the sum
\beq\label{riemann-spinor}
R_{\al(2)|\bt(2)}=W_{\al(2)\bt(2)}+\widehat R_{\al(2)|\bt(2)}.
\eeq
The first summand is the totally symmetric Weyl curvature spinor
\beq\label{weyl-spinor}
W_{\al(4)}=\frac13\left(R_{\al_1\al_2|\al_3\al_4}+R_{\al_1\al_3|\al_4\al_2}+R_{\al_1\al_4|\al_2\al_3}\right)
\eeq
and has 35 components, while the second
\beq\label{weyl-complement}
\widehat R_{\al(2)|\bt(2)}=\frac23\left(R_{\al_1\al_2|\bt_1\bt_2}-\frac12R_{\al_1\bt_1|\bt_2\al_2}-\frac12R_{\al_1\bt_2|\al_2\bt_1}\right)
\eeq
has the same symmetries as $R_{\al(2)|\bt(2)}$ but its symmetrized part vanishes leaving 20 independent components. $\wh R_{\al(2)|\bt(2)}$ can be further decomposed into three irreducible spinors
\beq\label{three-irrspinors}
\begin{array}{rl}
\widehat R_{\al(2)|\bt(2)}=&\bar R_{\al(2)|\bt(2)}-\frac16\left(C_{\al_1\bt_1}\wt R_{\al_2\bt_2}+C_{\al_1\bt_2}\wt R_{\al_2\bt_1}+C_{\al_2\bt_1}\wt R_{\al_1\bt_2}+C_{\al_2\bt_2}\wt R_{\al_1\bt_1}\right)\\[0.2cm]
+&\frac{1}{20}\widehat R\left(C_{\al_1\bt_1}C_{\al_2\bt_2}+C_{\al_1\bt_2}C_{\al_2\bt_1}\right).
\end{array}
\eeq
4-index irreducible curvature spinor $\bar R_{\al(2)|\bt(2)}$ has the same symmetries as $\widehat R_{\al(2)|\bt(2)}$ but additionally all possible traces that one can take using the charge conjugation matrix $C_{\g\de}$ vanish thus leaving 14 independent components. Note that defined in Ref.~\cite{Garcia} curvature spinor $\Omega_{\al_1\bt_1|\al_2\bt_2}$, that is antisymmetric in the first and the second pairs of indices, admits the following expression in terms of $\bar R_{\al(2)|\bt(2)}$
\beq
\Omega_{\al_1\bt_1|\al_2\bt_2}=\bar R_{\al_1\bt_2|\al_2\bt_1}-\bar R_{\al_1\al_2|\bt_1\bt_2}.
\eeq
2-index curvature spinor $\wt R_{\al[2]}$ being antisymmetric and traceless has 5 components, while the scalar $\hat R$ equals
\beq
\wh R=2R.
\eeq

Substitution of (\ref{three-irrspinors}) into (\ref{ricci-spinor-form}) allows to express spinor form of the Ricci tensor via the irreducible curvature spinors
\beq
\begin{array}{rl}
R_{\al[2]\bt[2]}=&\bar R_{\al_1\bt_2|\al_2\bt_1}-\bar R_{\al_1\bt_1|\al_2\bt_2}\\[0.2cm]
-&\frac{1}{10}\wh R\left(C_{\al_1\al_2}C_{\bt_1\bt_2}-2C_{\al_1\bt_1}C_{\al_2\bt_2}+2C_{\al_1\bt_2}C_{\al_2\bt_1}\right).
\end{array}
\eeq
The contribution of the 2-index curvature spinor $\wt R_{\al[2]}$ drops out because of the relation
\beq
C_{\bt\g}\wt R_{\de\varepsilon}+C_{\de\varepsilon}\wt R_{\bt\g}+C_{\g\de}\wt R_{\bt\varepsilon}+C_{\bt\varepsilon}\wt R_{\g\de}-C_{\bt\de}\wt R_{\g\varepsilon}-C_{\g\varepsilon}\wt R_{\bt\de}=0
\eeq
that can be derived from the identity
\beq
\wt R_{\lambda[\al}\varepsilon_{\bt\g\de\varepsilon]}=0
\eeq
upon taking into account the realization of the totally antisymmetric 4-index spinor in terms of the charge conjugation matrices
\beq
\varepsilon_{\al\bt\g\de}=-C_{\al\bt}C_{\g\de}+C_{\al\g}C_{\bt\de}-C_{\al\de}C_{\bt\g}
\eeq
and tracelessness of $\wt R_{\al[2]}$. In fact it can be shown that $\wt R_{\al[2]}$ vanishes since the algebraic Bianchi identity
\beq
R_{[klmn]}=0
\eeq
in the spinor form amounts to
\beq\label{algebraic-bianchi-spinor}
\wt R_{\al[2]}=0.
\eeq

From the definition of the Einstein tensor
\beq
\mc E_{mn}=R_{mn}-\frac12\eta_{mn}R
\eeq
it is possible to express it in terms of the irreducible curvature spinors
\beq
\mc E_{\al[2]\bt[2]}=\bar R_{\al_1\bt_2|\al_2\bt_1}-\bar R_{\al_1\bt_1|\al_2\bt_2}+\frac{3}{20}\wh R\left(C_{\al_1\al_2}C_{\bt_1\bt_2}-2C_{\al_1\bt_1}C_{\al_2\bt_2}+2C_{\al_1\bt_2}C_{\al_2\bt_1}\right).
\eeq
So that vacuum Einstein equations yield $\bar R_{\al(2)|\bt(2)}=\hat R=0$ leaving Weyl curvature spinor as the only non-vanishing quantity.

We conclude the discussion of $D=5$ Einstein gravity with the analysis of the spinor form of the second Bianchi identity
\beq
\mathscr D_{[k}R_{lm]pr}=0.
\eeq
Using the spinor form of the covariant derivative and the relation between the Riemann tensor and curvature spinor (\ref{rim-vec-spin}) one obtains equivalent form of the second Bianchi identity
\beq
B_{\al(2)|\bt(2)}=\mathscr D_{\al_1}\vp{\mathscr D}^{\lambda}R_{\lambda\al_2|\bt(2)}+\mathscr D_{\al_2}\vp{\mathscr D}^{\lambda}R_{\lambda\al_1|\bt(2)}=0.
\eeq
4-index spinor $B_{\al(2)|\bt(2)}$ is symmetric in the first and the second pairs of indices. It can be presented as the sum of two 4-index spinors symmetric and antisymmetric under permutation of the pairs of indices
\begin{subequations}
\begin{eqnarray}
B_{\al(2)|\bt(2)}&=&S_{\al(2)|\bt(2)}+A_{\al(2)|\bt(2)}:\\[0.2cm]
S_{\al(2)|\bt(2)}&=&\frac12\left(\mathscr D_{\al_1}\vp{\mathscr D}^{\lambda}R_{\lambda\al_2|\bt(2)}+\mathscr D_{\al_2}\vp{\mathscr D}^{\lambda}R_{\lambda\al_1|\bt(2)}+\mathscr D_{\bt_1}\vp{\mathscr D}^{\lambda}R_{\lambda\bt_2|\al(2)}+\mathscr D_{\bt_2}\vp{\mathscr D}^{\lambda}R_{\lambda\bt_1|\al(2)}\right),\label{second-bianchi-decomp}\\[0.2cm]
A_{\al(2)|\bt(2)}&=&\frac12\left(\mathscr D_{\al_1}\vp{\mathscr D}^{\lambda}R_{\lambda\al_2|\bt(2)}+\mathscr D_{\al_2}\vp{\mathscr D}^{\lambda}R_{\lambda\al_1|\bt(2)}-\mathscr D_{\bt_1}\vp{\mathscr D}^{\lambda}R_{\lambda\bt_2|\al(2)}-\mathscr D_{\bt_2}\vp{\mathscr D}^{\lambda}R_{\lambda\bt_1|\al(2)}\right).\label{second-bianchi-decomp2}
\end{eqnarray}
\end{subequations}
$S_{\al(2)|\bt(2)}$ has exactly the same symmetries as the Riemann curvature spinor. So its decomposition into irreducible spinors parallels that of $R_{\al(2)|\bt(2)}$ (cf. Eqs.~(\ref{riemann-spinor})-(\ref{three-irrspinors})). First one singles out totally symmetric part and that, whose symmetrization gives zero,
\beq
S_{\al(2)|\bt(2)}=S_{\al(2)\bt(2)}+\widehat S_{\al(2)|\bt(2)},
\eeq
where
\beq
S_{\al(4)}=\frac13\left(S_{\al_1\al_2|\al_3\al_4}+S_{\al_1\al_3|\al_4\al_2}+S_{\al_1\al_4|\al_2\al_3}\right)
\eeq
and
\beq
\widehat S_{\al(2)|\bt(2)}=\frac23\left(S_{\al_1\al_2|\bt_1\bt_2}-\frac12S_{\al_1\bt_1|\bt_2\al_2}-\frac12S_{\al_1\bt_2|\al_2\bt_1}\right).
\eeq
The totally symmetric part is contributed only by the Weyl curvature spinor (\ref{weyl-spinor})
\beq\label{S-symm}
S_{\al(4)}=\frac12\left(\mathscr D_{\al_1}\vp{\mathscr D}^{\lambda}W_{\lambda\al_2\al_3\al_4}+\mathscr D_{\al_2}\vp{\mathscr D}^{\lambda}W_{\lambda\al_3\al_4\al_1}+\mathscr D_{\al_3}\vp{\mathscr D}^{\lambda}W_{\lambda\al_4\al_1\al_2}+\mathscr D_{\al_4}\vp{\mathscr D}^{\lambda}W_{\lambda\al_1\al_2\al_3}\right),
\eeq
while $\widehat S_{\al(2)|\bt(2)}$ -- by $\widehat R_{\al(2)|\bt(2)}$ (\ref{weyl-complement})
\beq
\widehat S_{\al(2)|\bt(2)}=\frac12\left(\mathscr D_{\al_1}\vp{\mathscr D}^{\lambda}\widehat R_{\lambda\al_2|\bt(2)}+\mathscr D_{\al_2}\vp{\mathscr D}^{\lambda}\widehat R_{\lambda\al_1|\bt(2)}+\mathscr D_{\bt_1}\vp{\mathscr D}^{\lambda}\widehat R_{\lambda\bt_2|\al(2)}+\mathscr D_{\bt_2}\vp{\mathscr D}^{\lambda}\widehat R_{\lambda\bt_1|\al(2)}\right).
\eeq
For future reference let us adduce the decomposition of the covariant derivative of the Weyl curvature spinor
\beq\label{weyl-spinor-derivative}
\mathscr D_{\al}\vp{\mathscr D}^{\lambda}W_{\lambda\bt(3)}=\frac12S_{\al\bt(3)}+\mc W_{\al|\bt(3)},
\eeq
where
\beq
\mc W_{\al|\bt(3)}=\frac34\left(\mathscr D_{\al}\vp{\mathscr D}^{\lambda}W_{\lambda\bt_1\bt_2\bt_3}-\frac13\mathscr D_{\bt_1}\vp{\mathscr D}^{\lambda}W_{\lambda\bt_2\bt_3\al}-\frac13\mathscr D_{\bt_2}\vp{\mathscr D}^{\lambda}W_{\lambda\bt_3\al\bt_1}-\frac13\mathscr D_{\bt_3}\vp{\mathscr D}^{\lambda}W_{\lambda\al\bt_1\bt_2}\right)
\eeq
is symmetric in the last three indices but its symmetrization over all the four indices gives zero. $\widehat S_{\al(2)|\bt(2)}$ analogously to (\ref{three-irrspinors}) can be represented as the sum
\beq
\begin{array}{rl}
\widehat S_{\al(2)|\bt(2)}=&\bar S_{\al(2)|\bt(2)}-\frac16\left(C_{\al_1\bt_1}\wt S_{\al_2\bt_2}+C_{\al_1\bt_2}\wt S_{\al_2\bt_1}+C_{\al_2\bt_1}\wt S_{\al_1\bt_2}+C_{\al_2\bt_2}\wt S_{\al_1\bt_1}\right)\\[0.2cm]
+&\frac{1}{20}\wh S\left(C_{\al_1\bt_1}C_{\al_2\bt_2}+C_{\al_1\bt_2}C_{\al_2\bt_1}\right).
\end{array}
\eeq
Irreducible spinors on the r.h.s. have the same symmetries as corresponding curvature spinors on the r.h.s. of (\ref{three-irrspinors}). Explicit expressions for them via the covariant derivatives of the curvature spinors are found to be
\beq
\begin{array}{rl}
\bar S_{\al(2)|\bt(2)}=&\frac12\left(\mathscr D_{\al_1}\vp{\mathscr D}^{\lambda}\bar R_{\lambda\al_2|\bt(2)}+\mathscr D_{\al_2}\vp{\mathscr D}^{\lambda}\bar R_{\lambda\al_1|\bt(2)}+\mathscr D_{\bt_1}\vp{\mathscr D}^{\lambda}\bar R_{\lambda\bt_2|\al(2)}+\mathscr D_{\bt_2}\vp{\mathscr D}^{\lambda}\bar R_{\lambda\bt_1|\al(2)}\right)\\[0.2cm]
+&\!\frac{1}{12}\mathscr D^{\lambda\mu}\!\!\left(C_{\al_1\bt_1}\bar R_{\lambda\al_2|\mu\bt_2}\!+\! C_{\al_1\bt_2}\bar R_{\lambda\al_2|\mu\bt_1}\!+\! C_{\al_2\bt_1}\bar R_{\lambda\al_1|\mu\bt_2}\!+\! C_{\al_2\bt_2}\bar R_{\lambda\al_1|\mu\bt_1}\!\right)\!=\!0,\\[0.2cm]
\wt S_{\al[2]}=&\frac12\mathscr D^{\lambda\mu}\bar R_{\lambda\al_1|\mu\al_2}+\frac{3}{10}\mathscr D_{\al[2]}\widehat R=0,\\[0.2cm]
\widehat S=&0
\end{array}
\eeq
modulo terms proportional to the covariant derivative of the 2-index curvature spinor  $\mathscr D\wt R$ that vanish when the spinor form of the algebraic Bianchi identity (\ref{algebraic-bianchi-spinor}) is taken into account. Thus symmetric under permutation of the pairs of indices part of the second Bianchi identity (\ref{second-bianchi-decomp}) contains 54 non-trivial equations for the curvature spinors. For vacuum space-times there remain only 35 equations given by vanishing of (\ref{S-symm}).

Spinor $A_{\al(2)|\bt(2)}$ defined in (\ref{second-bianchi-decomp2}) can be decomposed into the traceless part and that contributing to the trace
\beq
A_{\al(2)|\bt(2)}=\bar A_{\al(2)|\bt(2)}-\frac16\left(C_{\al_1\bt_1}A_{\al_2\bt_2}+C_{\al_1\bt_2}A_{\al_2\bt_1}+C_{\al_2\bt_1}A_{\al_1\bt_2}+C_{\al_2\bt_2}A_{\al_1\bt_1}\right).
\eeq
Symmetric 2-index spinor $A_{\al(2)}$ vanishes, when the algebraic Bianchi identity is satisfied. For the traceless part we obtain
\beq
\begin{array}{rl}
\bar A_{\al(2)|\bt(2)}=&\frac12\left(\mc W_{\al_1|\al_2\bt_1\bt_2}+\mc W_{\al_2|\al_1\bt_1\bt_2}-\mc W_{\bt_1|\bt_2\al_1\al_2}-\mc W_{\bt_2|\bt_1\al_1\al_2}\right)\\[0.2cm]
+&\frac12\left(\mathscr D_{\al_1}\vp{\mathscr D}^{\lambda}\bar R_{\lambda\al_2|\bt(2)}+\mathscr D_{\al_2}\vp{\mathscr D}^{\lambda}\bar R_{\lambda\al_1|\bt(2)}-\mathscr D_{\bt_1}\vp{\mathscr D}^{\lambda}\bar R_{\lambda\bt_2|\al(2)}-\mathscr D_{\bt_2}\vp{\mathscr D}^{\lambda}\bar R_{\lambda\bt_1|\al(2)}\right)=0.
\end{array}
\eeq
For vacuum space-times it amounts to the vanishing of $\mc W_{\al|\bt(3)}$ so that recalling (\ref{weyl-spinor-derivative}) we come to the following equation for the Weyl curvature spinor
\beq
\mathscr D_{\al}\vp{\mathscr D}^{\lambda}W_{\lambda\bt(3)}=0
\eeq
and its linearization around flat background
\beq\label{weyl-spinor-eq}
\partial_{\al}\vp{\partial}^{\lambda}W_{\lambda\bt(3)}=0.
\eeq

Eq.~(\ref{weyl-spinor-eq}) is the spin-2 counterpart of Eqs.~(\ref{spin1eq}), (\ref{spin32eqn}) for spin-1 and spin-3/2 fields. In fact the sequence can be continued to accommodate (symmetric) higher-spin free fields. One can define a higher-spin generalization of the Weyl curvature spinor $W_{\al(2s)}$ that should satisfy the equation generalizing (\ref{weyl-spinor-eq})
\beq\label{hs-spinor-eq}
\partial_{\al}\vp{\partial}^{\lambda}W_{\lambda\bt(2s-1)}=0.
\eeq
This equation encompasses both low- and higher-spin cases and
applies to arbitrary (half-)integer $s\geq1/2$. It can be shown \cite{Lopatin} that for higher-spin fields on-shell there remains non-zero only the Weyl tensor similarly to the low-spin ones.

\section{On-shell integral representation for Weyl curvature spinors of $D=5$ massless gauge fields}
\label{harmonic-section}

Equation (\ref{hs-spinor-eq}) can be solved using the integral formula that maps functions $\phi_{i(2s)}$ on $S^3$ that carry $2s$ symmetrized indices of the fundamental representation of $SU(2)$ (and possibly indices of the $R-$symmetry group representations) to Weyl curvature spinors of spin-$s$ fields in Minkowski space. The role of mediator is played by the $D=5$ Lorentz-harmonic spinor variables. Present section aims at giving the details of such a construction being the generalization of that elaborated in Ref.~\cite{DelducGS} for the string-theoretic dimensions $D=3,4,6,10$.\footnote{In \cite{DelducGS} there was also given on-shell integral representation for integer-spin fields based on $D-$dimensional vector harmonics.} To this end we start by recapitulating necessary properties of $D=5$ Lorentz-harmonic variables.

Vector Lorentz harmonics are given by $5\times5$ matrix $n_{m}\vp{n}^{\mathbf{m}}$ subject to the constraint
\beq\label{vector-orthnorm}
n_{m}\vp{n}^{\mathbf{m}}\eta^{mn}n_{n}\vp{n}^{\mathbf{n}}=\eta^{\mathbf{m}\mathbf{n}},\quad\eta=\mr{diag}(-,+,+,+,+)
\eeq
so that it takes value in the Lorentz group $SO(1,4)$. Light-face indices are acted upon by the left $SO(1,4)_L$ rotations, while the bold-face ones transform under the right $SO(1,4)_R$ group, whose $SO(1,1)\times ISO(3)$ subgroup will be gauged. Taking the first and, e.g., the last columns of the harmonic matrix allows to define two light-like vectors
\beq\label{null-vectors}
n_{m}\vp{n}^{\mathbf{\pm2}}=n_{m}\vp{n}^{\mathbf0}\pm n_{m}\vp{n}^{\mathbf5}:\quad n_{m}\vp{n}^{\mathbf{\pm2}}n^{m\mathbf{\pm2}}=0,\quad n_{m}\vp{n}^{\mathbf{\pm2}}n^{m\mathbf{\mp2}}=-2.
\eeq
This introduces decomposition of the vector harmonic matrix into three blocks
\beq\label{vect-harm-decomp}
n_{m}\vp{n}^{\mathbf{m}}=(n_{m}\vp{n}^{\mathbf{\pm2}}, n_{m}\vp{n}^{\mathbf{I}}),\quad\mathbf I=1,2,3
\eeq
reducing manifest $SO(1,4)_R$-covariance down to $SO(1,1)\times SO(3)$. Above introduced components of the matrix $n_{m}\vp{n}^{\mathbf{m}}$ transform in the following manner under infinitesimal $SO(1,4)_R$ rotation with parameters $L^{\mathbf{m}\mathbf{n}}=(L^{\mathbf{+2}\mathbf{-2}}, L^{\mathbf{\pm2}\mathbf{I}}, L^{\mathbf{I}\mathbf{J}})$
\beq\label{so-r-inf}
\begin{array}{c}
\delta n_{m}\vp{n}^{\mathbf{\pm2}}=\pm L^{\mathbf{+2-2}}n_{m}\vp{n}^{\mathbf{\pm2}}+L^{\mathbf{\pm2 I}}n_{m}\vp{n}^{\mathbf{I}},\\[0.2cm]
\delta n_{m}\vp{n}^{\mathbf{I}}=-\frac12\left(L^{\mathbf{+2 I}}n_{m}\vp{n}^{\mathbf{-2}}+L^{\mathbf{-2 I}}n_{m}\vp{n}^{\mathbf{+2}}\right)+L^{\mathbf{IJ}}n_{m}\vp{n}^{\mathbf{J}}.
\end{array}
\eeq
Due to zero norm any of the vectors (\ref{null-vectors}) can be set proportional to $D=5$ massless particle's momentum. For instance, $n_{m}\vp{n}^{\mathbf{+2}}$ is invariant under the transformations with parameters $L^{\mathbf{IJ}}$ and $L^{\mathbf{-2I}}$, and covariant under those with the parameter $L^{\mathbf{+2-2}}$ in (\ref{so-r-inf}) corresponding to the $SO(1,1)\times ISO(3)$ subgroup of $SO(1,4)_R$. This subgroup can be gauged, so that the vector harmonic matrix will parametrize the $SO(1,4)/SO(1,1)\times ISO(3)$ coset-space isomorphic to the 3-sphere. More explicitly one can write $n_{m}\vp{n}^{\mathbf{+2}}=(q^{+2}, q^{+2}k_{\hat m})$ with the Euclidean 4-vector $k_{\hat m}$: $k^2=1$ parametrizing $S^3$.

Like in arbitrary dimension, $SO(1,4)-$valued matrix can be presented as the 'square' of the $Spin(1,4)$ matrix $v^{\al\boldsymbol{\mu}}$
\beq
\label{vector-vs-spinor-harmonics}
n_{m}\vp{n}^{\mathbf{m}}=\frac14 v^{\al\boldsymbol{\mu}}\g_{m\al}\vp{\g}^{\bt}\g^{\mathbf{m}}\vp{\g}_{\boldsymbol{\mu}}\vp{\g}^{\boldsymbol{\nu}}v_{\bt\boldsymbol{\nu}},\quad v_{\bt\boldsymbol{\nu}}=C_{\bt\g}C_{\boldsymbol{\nu\lambda}}v^{\g\boldsymbol{\lambda}}.
\eeq
In analogy with the vector-harmonic matrix, the light-face index of $v^{\al\boldsymbol{\mu}}$ transforms under the spinor representation of $SO(1,4)_L$ and the bold-face index under that of $SO(1,4)_R$. Orthonormality conditions (\ref{vector-orthnorm}) are then satisfied by virtue of six harmonicity conditions imposed on the $4\times4$ spinor-harmonic matrix
\beq\label{harm-cond}
v^{\al\boldsymbol{\mu}}C_{\al\bt}v^{\bt\boldsymbol{\nu}}=C^{\boldsymbol{\mu\nu}},
\eeq
where $C_{\al\bt}$ and $C^{\boldsymbol{\mu\nu}}$ are charge conjugation matrices acting on the $SO(1,4)_L$ and $SO(1,4)_R$ spinor indices respectively. Harmonicity conditions ensure that $v^{\al\boldsymbol{\mu}}$ takes value in $Spin(1,4)$. Similarly to the decomposition (\ref{vect-harm-decomp}) of the vector harmonic matrix on $SO(1,1)\times SO(3)$ covariant blocks spinor harmonics decompose into the following $4\times 2$ blocks
\beq\label{spinor-harmonics}
v^{\al\boldsymbol{\mu}}=
\left(
\begin{array}{c}
v^{\al+i}\\ v^{\al-i}
\end{array}
\right),
\eeq
where $i$ is the $SU(2)$ fundamental representation index.
In terms of these $4\times 2$ blocks null components of the vector harmonics $n_{m}\vp{n}^{\mathbf{\pm2}}$ acquire the form
\beq
n_{m}\vp{n}^{\mathbf{\pm2}}=\frac12v^{\al\pm i}\g_{m\al}\vp{\g}^{\bt}v_{\bt}\vp{v}^{\pm}\vp{v}_i.
\eeq
Fulfilment of the zero norm and orthonormality conditions (\ref{null-vectors}) is again by virtue of the harmonicity conditions (\ref{harm-cond}), whose component form is
\beq\label{harmcond-component}
v^{\al\,\pm i}v_{\al}\vp{v}^{\pm j}=0,\quad v^{\al\,\pm i}v_{\al}\vp{v}^{\mp j}-i\varepsilon^{ij}=0.
\eeq
More details on the properties of $D=5$ spinor harmonics can be found in Appendix B.

Harmonic variables can be used to construct $SO(1,4)_L-$invariant Cartan 1-forms. Those taking value in the Lie algebra of $SO(1,4)_R$ are defined by the relations
\beq
\Omega^{\boldsymbol{mn}}(d)=\frac12\left(n_{m}\vp{n}^{\boldsymbol{n}}dn^{m\boldsymbol{m}}-n_{m}\vp{n}^{\boldsymbol{m}}dn^{m\boldsymbol{n}}\right)=\frac12v^{\al\boldsymbol{\mu}}\g^{\boldsymbol{mn}}\vp{\g}_{\boldsymbol{\mu}}\vp{\g}^{\boldsymbol{\nu}}dv_{\al\boldsymbol{\nu}}.
\eeq
They further decompose on 4 irreducible components under the $SO(1,1)\times SO(3)$ subgroup of $SO(1,4)_R$. Explicit expressions for them in terms of the spinor harmonics will be used below
\beq\label{cartan-forms-definition}
\begin{array}{rl}
\Omega^{\boldsymbol{+2-2}}(d)=&i(v^{\al+i}dv_\al\vp{v}^-\vp{v}_i-v^{\al-i}dv_\al\vp{v}^+\vp{v}_i),\\[0.2cm]
\Omega^{\boldsymbol{\pm2I}}(d)=&-v^{\al\pm i}\tau^{I}\vp{\tau}_i\vp{\tau}^jdv_{\al}\vp{v}^{\pm}\vp{v}_j,\\[0.2cm]
\Omega^{\boldsymbol{IJ}}(d)=&\frac12\varepsilon^{IJK}(v^{\al+i}\tau^{K}\vp{\tau}_i\vp{\tau}^jdv_{\al}\vp{v}^{-}\vp{v}_j+v^{\al-i}\tau^{K}\vp{\tau}_i\vp{\tau}^jdv_{\al}\vp{v}^{+}\vp{v}_j).
\end{array}
\eeq

Proceed now to discussion of the integral representation for the
Weyl curvature spinors of the gauge fields. It has the following
form
\beq\label{integral}
W_{\al(2s)}(x^{m})=\int\limits_{S^3}\Omega^{+6}v_{\al_1}\vp{v}^{+i_1}\cdots
v_{\al_{2s}}\vp{v}^{+i_{2s}}\phi^{-6-2s}\vp{\phi}_{i(2s)}(x^{+2},v^+).
\eeq
Harmonic measure is given by the 3-form
\beq
\Omega^{+6}=\varepsilon^{IJK}\Omega^{\boldsymbol{+2I}}\wedge\Omega^{\boldsymbol{+2J}}\wedge\Omega^{\boldsymbol{+2K}}.
\eeq
It is $ISO(3)$ invariant (corresponding parameters in
(\ref{so-r-inf}) are $L^{\boldsymbol{-2I}}$ and
$L^{\boldsymbol{IJ}}$) and $SO(1,1)$ covariant so that the
integral is $SO(1,1)\times ISO(3)$ invariant. By appropriate
parametrization of harmonics it reduces to the standard measure on
$S^3$ written via angle variables. Integrand
\beq\label{integrand}
\phi^{-6-2s}\vp{\phi}_{i(2s)}(x^{+2},v^+)
\eeq
transforms
homogeneously under $SO(1,1)$ with weight $-6-2s$ to compensate
contributions of harmonic measure and spinor harmonics and depends
on the space-time coordinates only through the projection
$x^{+2}=x^{m}n_{m}\vp{n}^{\boldsymbol{+2}}$. This ensures that
$W_{\al(2s)}$ defined by (\ref{integral}) satisfies
Eq.~(\ref{hs-spinor-eq}) for $s>0$ and $\Box W_{\al(2s)}=0$ for
$s\geq0$. Note that the integral representation analogous to that
of Eq.~(\ref{integral}) can also be constructed using
$v_{\al}\vp{v}^{-i}$ spinor. In that case homogeneity degree in
$v_{\al}\vp{v}^{-i}$ of the integrand should be opposite to that
in (\ref{integrand}).

\section{Massless particle model in doubly harmonic superspace}

In this section we discuss a toy model of the massless
superparticle, whose canonical quantization yields functions
(\ref{integrand}) with various values of $s$ assembled into
multiplets of $D=5$ Poincare supersymmetry. It is characterized by
the following action\footnote{In this section, because we
introduce harmonics $w^{0,\pm i}$ parametrizing the two-sphere
$SU(2)/U(1)$ in addition to the Lorentz harmonics, the notation
has to be slightly improved. For any quantity $\mc Q^{\al p,rT}$
Lorentz indices (if any) precede integers $p$ and $r$ separated by
comma that denote $SO(1,1)$ and $U(1)$ weights respectively. They
are followed by other indices that in the above example have been
collectively denoted by $T$. Thus for the introduced in the
previous section spinor Lorentz harmonics (\ref{spinor-harmonics})
and $SO(1,4)_L-$invariant Cartan forms
(\ref{cartan-forms-definition}) it is additionally specified that
they carry zero weight w.r.t. to the $U(1)$ subgroup of $SU(2)$,
namely $v^{\al\pm i}\equiv v^{\al\pm,0i}$,
$\Omega^{\boldsymbol{+2-2}}(d)\equiv\Omega^{+2-2,0}(d)$,
$\Omega^{\boldsymbol{\pm2I}}(d)\equiv\Omega^{+2,0I}(d)$ etc.}
\beq\label{superparticle-action}
\begin{array}{rl}
S=&\int d\tau\mathscr L,\\[0.2cm]
\mathscr L=&p^{-2,0}\dot x^{+2,0}+\pi^{+,-a}\dot\theta^{-,+}\vp{\theta}_{a}+\frac12\Omega^{+2-2,0}_\tau(p^{-2,0}x^{+2,0}+\frac12\pi^{+,-a}\theta^{-,+}\vp{\theta}_{a})\\[0.2cm]
+&\Omega^{+2,0I}_\tau y^{-2,0I}+\omega^{0,0}_\tau\pi^{+,- a}\theta^{-,+}\vp{\theta}_{a}+\omega^{0,+2}_\tau y^{0,-2}\\[0.2cm]
+&q\Omega^{+2-2,0}_\tau+\ell\omega^{0,0}_\tau.
\end{array}
\eeq
Coordinate $x^{+2,0}$ (denoted as $x^{+2}$ in section
\ref{harmonic-section}) upon quantization will become the argument
of  the superparticle's wave-function (cf. (\ref{integrand})),
Grassmann odd coordinate $\theta^{-,+}\vp{\theta}_{a}$ in addition
to $SO(1,1)$ and $U(1)$ weights of unit modulus carries index
$a=1,\ldots,N$ labeling $D=5$ supersymmetries; $p^{-2,0}$ and
$\pi^{+,-a}$ are their canonical momenta. Action
(\ref{superparticle-action}) also contains four non-dynamic
bosonic coordinates $y^{-2,0I}$ and $y^{0,-2}$ and two sets of
harmonic variables. $Spin(1,4)$ Lorentz harmonics
$v^{\al\pm,\,0i}$ enter via the world-line projections of Cartan
forms $\Omega^{+2-2,0}_\tau$ and $\Omega^{+2,0I}_\tau$ (cf.
(\ref{cartan-forms-definition})). Due to the $ISO(3)$ gauge
symmetry with parameters $L^{0,0IJ}(\tau)$ and
$L^{-2,0I}(\tau)$, under which variables entering the
superparticle's Lagrangian transform as
\beq\label{so3-gauge-sym}
\de\Omega^{+2-2,0}_\tau=0,\quad\de\Omega^{+2,0I}_\tau=L^{0,0IJ}\Omega^{+2,0J}_\tau,\quad\de
y^{-2,0I}=L^{0,0IJ}y^{-2,0J}
\eeq
and
\beq\label{3-shift-gauge-sym}
\begin{array}{c}
\de\Omega^{+2-2,0}_\tau=L^{-2,0I}\Omega^{+2,0I}_\tau,\quad\de\Omega^{+2,0I}_\tau=0,\\[0.2cm]
\de y^{-2,0I}=-\frac12L^{-2,0I}\left(p^{-2,0}x^{+2,0}+\frac12\pi^{+,-a}\theta^{-,+}\vp{\theta}_{a}+2q\right),
\end{array}
\eeq
Lorentz harmonics parametrize the $SO(1,4)/ISO(3)$
manifold. The action also depends on another set of harmonics
$w^{0,\pm i}$ via the world-line projections of the Cartan forms
\beq\label{su-cartan-forms-def} \omega^{0,0}_\tau=\frac12(\dot
w^{0,+i}w^{0,-}\vp{w}_i+\dot
w^{0,-i}w^{0,+}\vp{w}_i),\quad\omega^{0,+2}_\tau=\dot
w^{0,+i}w^{0,+}\vp{w}_i.
\eeq
They are subject to the harmonicity condition
\beq\label{su2-harmcond}
w^{0,+i}w^{0,-}\vp{w}_i-1=0
\eeq
that is nothing but the
unimodularity condition ensuring that harmonic matrix $(w^{0,+i},w^{0,-i})$ takes value in
the $SU(2)$ group. Due to the gauge symmetry of the action
(\ref{superparticle-action}) with the parameter
$\Lambda^{0,-2}(\tau)$
\beq\label{0-2-gauge-sym}
\de\omega^{0,0}_\tau=\Lambda^{0,-2}\omega^{0,+2}_\tau,\quad\de\omega^{0,+2}_\tau=0,\quad\de
y^{0,-2}=-\Lambda^{0,-2}(\pi^{+,-a}\theta^{-,+}\vp{\theta}_{a}+\ell),
\eeq
these harmonics parametrize $S^2$ manifold. Two last summands
in (\ref{superparticle-action}) are harmonic $1d$ Wess-Zumino
terms. The first depends on the Lorentz harmonics and the
second -- on the $SU(2)/U(1)$ harmonics.\footnote{Superparticle
models with Wess-Zumino terms constructed out the $D=4$ spinor
harmonics were considered in \cite{Bandos-JETP}, \cite{Bandos'90}
and those constructed out of the $SU(2)$ harmonics in
\cite{Akulov'88}.} Values of the numerical coefficients $q$ and
$\ell$ will be determined below in terms of $N$ and the maximal
spin in the supermultiplet.

Systematic treatment of gauge symmetries is achieved in the framework of the canonical approach to the discussion of which we now turn. Definition of canonical momenta results in the following primary constraints
\begin{subequations}
\begin{eqnarray}
P^{+2,0I}&\approx&0,\label{zero-momentum1}\\[0.2cm]
P^{0,+2}&\approx&0;\label{zero-momentum2}\\[0.2cm]
T_{\al}\vp{T}^{-,0}\vp{T}_{i}&=&P_{\al}\vp{P}^{-,0}\vp{P}_{i}\!+\!\frac{i}{2}(p^{-2,0}x^{+2,0}\!+\!\frac12\pi^{+,-a}\theta^{-,+}\vp{\theta}_{a}\!+\!2q)v_{\al}\vp{v}^{-,0}\vp{v}_i\!-\! y^{-2,0I}\tau^I\vp{\tau}_i\vp{\tau}^jv_{\al}\vp{v}^{+,0}\vp{v}_j\approx0,\label{lorharm-momenta1}\\[0.2cm]
T_{\al}\vp{T}^{+,0}\vp{T}_{i}&=&P_{\al}\vp{P}^{+,0}\vp{P}_{i}-\frac{i}{2}(p^{-2,0}x^{+2,0}+\frac12\pi^{+,-a}\theta^{-,+}\vp{\theta}_{a}+2q)v_{\al}\vp{v}^{+,0}\vp{v}_i\approx0;\label{lorharm-momenta2}\\[0.2cm]
U^{0,-}\vp{U}_i&=&P^{0,-}\vp{P}_i-\frac12(\pi^{+,-a}\theta^{-,+}\vp{\theta}_{a}+\ell)w^{0,-}\vp{w}_i-y^{0,-2}w^{0,+}\vp{w}_i\approx0,\label{harm-momenta1}\\[0.2cm]
U^{0,+}\vp{U}_i&=&P^{0,+}\vp{P}_i-\frac12(\pi^{+,-a}\theta^{-,+}\vp{\theta}_{a}+\ell)w^{0,+}\vp{w}_i\approx0, \label{harm-momenta2}
\end{eqnarray}
\end{subequations}
where $P^{+2,0I}$ and $P^{0,+2}$ are momenta conjugate to
$y^{-2,0I}$ and $y^{0,-2}$, $P_{\al}\vp{P}^{\mp,0}\vp{P}_{i}$ are
momenta for the Lorentz harmonics $v^{\al\pm,0i}$ and $P^{0,\mp}\vp{P}_i$ -- for
the $SU(2)$ harmonics $w^{0,\pm i}$. This set of constraints has to be
supplemented by the harmonicity conditions
(\ref{harmcond-component}) and (\ref{su2-harmcond}) that in the
canonical approach should be treated as weak equalities in the
sense of Dirac. Since it is convenient to consider them holding as
usual equalities, i.e. in the strong sense according to Dirac, the
following technical trick can be applied. It is possible to single
out part of the constraints on the harmonic momenta that form with
the harmonicity conditions conjugate pairs of the second-class
constraints and construct associated Dirac brackets. Remaining
constraints on the harmonic momenta take value in the right
Lorentz algebra and are called covariant momenta. Pivotal property
of such Dirac brackets is that on the subspace of the harmonic
phase-space spanned by harmonics themselves and covariant momenta
they coincide with the Poisson brackets, while remaining
constraints including the harmonicity conditions are then
fulfilled in the strong sense.\footnote{For detailed discussion of
the Hamiltonian description of the Lorentz harmonics see, e.g.
\cite{BZ-ICTP'92}.}

Consider in more detail $4+1$ constraints (\ref{harm-momenta1}), (\ref{harm-momenta2}) and (\ref{su2-harmcond}) in the sector of $SU(2)$ harmonics. Projecting constraints (\ref{harm-momenta1}) and (\ref{harm-momenta2}) onto the harmonics and taking their linear combinations it is easy to find that the constraint
\beq
w^{0,+i}U^{0,-}\vp{U}_i+w^{0,-i}U^{0,+}\vp{U}_i=w^{0,+i}P^{0,-}\vp{P}_i+w^{0,-i}P^{0,+}\vp{P}_i\approx0
\eeq
forms with the harmonicity condition (\ref{su2-harmcond}) the pair of the second-class constraints that can be converted into strong equalities by introducing Dirac brackets. So that in the sector of $SU(2)$ harmonics there remain three constraints corresponding to covariant momenta generating the $su(2)_R$ algebra
\begin{subequations}
\begin{eqnarray}
\mathscr R^{\:0,0}&=&w^{0,+i}U^{0,-}\vp{U}_i\!-\!w^{0,-i}U^{0,+}\vp{U}_i=w^{0,+i}P^{0,-}\vp{P}_i\!-\!w^{0,-i}P^{0,+}\vp{P}_i\!+\!\theta^{-,+}\vp{\theta}_{a}\pi^{+,-a}\!-\!\ell\approx0,\label{r00}\\[0.2cm]
\mathscr R^{\:0,+2}&=&w^{0,+i}U^{0,+}\vp{U}_i=w^{0,+i}P^{0,+}\vp{P}_i\approx0,\label{r0+2}\\[0.2cm]
\mathscr R^{\:0,-2}&=&w^{0,-i}U^{0,-}\vp{U}_i=w^{0,-i}P^{0,-}\vp{P}_i+y^{0,-2}\approx0.\label{r0-2}
\end{eqnarray}
\end{subequations}
In view of the constraints (\ref{zero-momentum2}) and (\ref{r0-2}) canonical pair $(P^{0,+2}, y^{0,-2})$ can be excluded from the consideration and in the sector of $SU(2)$ harmonics there remain just two constraints (\ref{r00}) and (\ref{r0+2}). The latter is the generator of the gauge symmetry (\ref{0-2-gauge-sym}).

In the sector of Lorentz harmonics there are 16+6 constraints (\ref{lorharm-momenta1}), (\ref{lorharm-momenta2}) and (\ref{harmcond-component}). Like in the case of $SU(2)$ harmonics, one can single out six $SO(1,4)_L-$invariant constraints from (\ref{lorharm-momenta1}), (\ref{lorharm-momenta2}) that form conjugate pairs of the second-class constraints with the harmonicity conditions (\ref{harmcond-component}). 10 remaining constraints constitute covariant momenta taking value in the $so(1,4)_R$ algebra
\begin{subequations}
\begin{eqnarray}
\mathscr P^{+2-2,0}&=&2(v^{\al-,0i}T_{\al}\vp{T}^{+,0}\vp{T}_{i}-v^{\al+,0i}T_{\al}\vp{T}^{-,0}\vp{T}_{i})\nonumber\\[0.2cm]
&=&-2(v^{\al+,0i}P_{\al}\vp{P}^{-,0}\vp{P}_{i}-v^{\al-,0i}P_{\al}\vp{P}^{+,0}\vp{P}_{i}\label{p+2-2}\\[0.2cm]
&+&2x^{+2,0}p^{-2,0}-\theta^{-,+}\vp{\theta}_{a}\pi^{+,-a}+4q)\approx0,\nonumber\\[0.2cm]
\mathscr P^{+2,0I}&=&-2iv^{\al+,0i}\tau^I\vp{\tau}_i\vp{\tau}^jT_{\al}\vp{T}^{+,0}\vp{T}_{j}=-2iv^{\al+,0i}\tau^I\vp{\tau}_i\vp{\tau}^jP_{\al}\vp{P}^{+,0}\vp{P}_{j}\approx0,\label{p+20I}\\[0.2cm]
\mathscr P^{-2,0I}&=&2iv^{\al-,0i}\tau^I\vp{\tau}_i\vp{\tau}^jT_{\al}\vp{T}^{-,0}\vp{T}_{j}=2iv^{\al-,0i}\tau^I\vp{\tau}_i\vp{\tau}^jP_{\al}\vp{P}^{-,0}\vp{P}_{j}-4y^{-2,0I}\approx0,\label{p-20I}\\[0.2cm]
\mathscr P^{\:0,0IJ}&=&i\varepsilon^{IJK}(v^{\al+,0i}\tau^K\vp{\tau}_i\vp{\tau}^jT_{\al}\vp{T}^{-,0}\vp{T}_{j}+v^{\al-,0i}\tau^K\vp{\tau}_i\vp{\tau}^jT_{\al}\vp{T}^{+,0}\vp{T}_{j})\label{pIJ}\\[0.2cm]
&=&i\varepsilon^{IJK}(v^{\al+,0i}\tau^K\vp{\tau}_i\vp{\tau}^jP_{\al}\vp{P}^{-,0}\vp{P}_{j}+v^{\al-,0i}\tau^K\vp{\tau}_i\vp{\tau}^jP_{\al}\vp{P}^{+,0}\vp{P}_{j})\approx0.\nonumber
\end{eqnarray}
\end{subequations}
One observes that due to the constraints (\ref{zero-momentum1}) and (\ref{p-20I}) canonical variables $(P^{+2,0I}, y^{-2,0I})$ can be excluded from consideration. So that in the sector of Lorentz harmonics one is left with seven constraints (\ref{p+2-2}), (\ref{p+20I}), (\ref{pIJ}). Constraints (\ref{pIJ}) and (\ref{p+20I}) are the generators of the discussed above gauge transformations (\ref{so3-gauge-sym}) and (\ref{3-shift-gauge-sym}) respectively.

Thus the constraints (\ref{r00}), (\ref{r0+2}) and (\ref{p+2-2}),
(\ref{p+20I}), (\ref{pIJ}) form the set of the first-class
constraints of our model. Associated quantum operators are imposed
on the superparticle's wave function that in the coordinate
representation depends on $x^{+2,0}$, $v^{\al\pm,0i}$, $w^{0,\pm
i}$ and $\theta^{-,+}\vp{\theta}_{a}$. Conjugate momenta then act
as differential operators \beq \hat
p^{-2,0}=\frac{\partial}{\partial x^{+2,0}},\quad\hat
P_{\al}\vp{P}^{\pm,0}\vp{P}_{i}=\frac{\partial}{\partial
v^{\al\mp,0i}},\quad \hat
P^{0,\pm}\vp{P}_i=\frac{\partial}{\partial w^{0,\mp
i}},\quad\hat\pi^{+,-a}=\frac{\partial}{\partial\theta^{-,+}\vp{\theta}_{a}}
\eeq so that the (anti)commutations relations hold \beq [\hat
p^{-2,0},\hat x^{+2,0}]=1,\quad [\hat
P_{\al}\vp{P}^{\pm,0}\vp{P}_{i},\hat
v^{\bt\mp,0j}]=\de_{\al}^{\bt}\de_i^j,\quad [\hat
P^{0,\pm}\vp{P}_i,\hat w^{0,\mp
j}]=\de_i^j,\quad\{\hat\pi^{+,-a},\hat\theta^{-,+}\vp{\theta}_{b}\}=\de^a_b.
\eeq
Classical generator of the $SO(1,1)$ gauge symmetry (\ref{p+2-2}) in
such a realization transforms into the following Hermitian
operator
\beq\label{so1,1q}
\widehat\Delta_{so(1,1)}=2x^{+2,0}\frac{\partial}{\partial
x^{+2,0}}+v^{\al+,0i}\frac{\partial}{\partial
v^{\al+,0i}}-v^{\al-,0i}\frac{\partial}{\partial
v^{\al-,0i}}-\theta^{-,+}\vp{\theta}_{a}\frac{\partial}{\partial\theta^{-,+}\vp{\theta}_{a}}-c_{so(1,1)},
\eeq where $c_{so(1,1)}=-1-4q-\frac{N}{2}$ is an ordering constant
. It is nothing but the dilatation operator acting on the
variables with non-zero $SO(1,1)$ weights. Similarly quantum
$U(1)$ generator associated with the constraint (\ref{r00}) can be
brought to the form \beq\label{u1q}
\widehat\Delta_{u(1)}=w^{0,+i}\frac{\partial}{\partial
w^{0,+i}}-w^{0,-i}\frac{\partial}{\partial
w^{0,-i}}+\theta^{-,+}\vp{\theta}_{a}\frac{\partial}{\partial\theta^{-,+}\vp{\theta}_{a}}-c_{u(1)},
\eeq where $c_{u(1)}=\ell+\frac{N}{2}$. It 'measures' the $U(1)$
charges of the components of superparticle's wave function.
Quantum operators corresponding to other constraints
(\ref{p+20I}), (\ref{pIJ}) and (\ref{r0+2}) are free from the
ordering ambiguities and can be defined in the following way
\begin{subequations}
\begin{eqnarray}
\hat{\mathrm P}^{+2,0I}&=&v^{\al+,0i}\tau^I\vp{\tau}_i\vp{\tau}^j\frac{\partial}{\partial v^{\al-,0j}}.\label{q+20I}\\[0.2cm]
\hat{\mathrm P}^{0,0IJ}&=&\varepsilon^{IJK}\left(v^{\al+,0i}\tau^{K}\vp{\tau}_i\vp{\tau}^j\frac{\partial}{\partial v^{\al+,0j}}+v^{\al-,0i}\tau^K\vp{\tau}_i\vp{\tau}^j\frac{\partial}{\partial v^{\al-,0j}}\right),\label{qIJ}\\[0.2cm]
\hat{\mathrm R}^{0,+2}&=&w^{0,+i}\frac{\partial}{\partial w^{0,-i}}.\label{q0+2}
\end{eqnarray}
\end{subequations}
These operators act on the wave function that admits series expansion in $\theta$
\beq\label{theta-expansion}
\begin{array}{rl}
\Phi^{c_{so(1,1)},\, c_{u(1)}}(x^{+2,0},v^{\al\pm,0i},w^{0,\pm i},\theta^{-,+}\vp{\theta}_{a})=&\varphi^{c_{so(1,1)},\, c_{u(1)}}+\theta^{-,+}\vp{\theta}_{a}\varphi^{c_{so(1,1)}+1,\, c_{u(1)}-1\: a}\\[0.2cm]
+&\cdots+\theta^{-k,+k}\vp{\theta}_{a[k]}\varphi^{c_{so(1,1)}+k,\, c_{u(1)}-k\, a[k]}\\[0.2cm]
+&\cdots+\theta^{-N,+N}\vp{\theta}_{a[N]}\varphi^{c_{so(1,1)}+N,\, c_{u(1)}-N\, a[N]},
\end{array}
\eeq
where $\theta^{-k,+k}\vp{\theta}_{a[k]}\equiv\theta^{-,+}\vp{\theta}_{a_1}\theta^{-,+}\vp{\theta}_{a_2}\cdots\theta^{-,+}\vp{\theta}_{a_k}$ and the component functions depend on $x^{+2,0}$ and harmonics. In particular, one finds that for any $0\leq k\leq N$
\begin{subequations}
\begin{eqnarray}
&\hat{\mathrm P}^{+2,0I}\varphi^{c_{so(1,1)}+k,\, c_{u(1)}-k\, a[k]}=0,&\\[0.2cm]
&\hat{\mathrm R}^{0,+2}\varphi^{c_{so(1,1)}+k,\, c_{u(1)}-k\, a[k]}=0&\label{r0+2phi}
\end{eqnarray}
\end{subequations}
and the operator (\ref{qIJ}) determines how the wave function components transform under the $SO(3)\sim SU(2)$.
The first equation implies that the wave function is independent of $v^{\al-,0i}$ (cf. \cite{DelducGS}) and the second allows to factorize the dependence on the $SU(2)$ harmonics. Namely, for any function with non-negative value of the $U(1)$ weight $c_{u(1)}-k\geq0$ it was proved in \cite{GIKOS} that the solution of Eq.~(\ref{r0+2phi}) is
\beq
\varphi^{c_{so(1,1)}+k,\, c_{u(1)}-k\, a[k]}(x^{+2,0},v^{\al+,0i},w^{0,+i})=\varphi^{c_{so(1,1)}+k,\,0\, a[k]}\vp{varphi}_{i(c_{u(1)}-k)}w^{0,+i(c_{u(1)}-k)},
\eeq
where $w^{0,+i(c_{u(1)}-k)}\equiv w^{0,+i_1}w^{0,+i_2}\cdots w^{0,+i_{c_{u(1)}-k}}$, while if the $U(1)$ weight is negative the function $\varphi^{c_{so(1,1)}+k,\, c_{u(1)}-k\, a[k]}$ vanishes. As a result component functions $\varphi^{c_{so(1,1)}+k,\,0\, a[k]}\vp{varphi}_{i(c_{u(1)}-k)}$ depend on $x^{+2,0}$ and $v^{\al+,0i}$ only. In order to identify them with the integrand in (\ref{integral}) the following relation between the ordering constants in (\ref{so1,1q}) and (\ref{u1q}) should hold
\beq
c_{so(1,1)}=-6-c_{u(1)}
\eeq
and then $c_{u(1)}-k=2s_k$, where $s_k$ is the spin of the corresponding gauge field. Using that the field with the highest value of spin in the multiplet $s_{max}$ corresponds to the leading component in the expansion (\ref{theta-expansion}), allows to express coefficients at the Wess-Zumino terms $q$ and $\ell$ through $s_{max}$.

Consider in more detail the case of $N=2$ supersymmetry. From equations
\beq
c_{so(1,1)}=-6-2s_{max},\quad c_{u(1)}=2s_{max}
\eeq
it follows that
\beq
q=1+\frac{s_{max}}{2},\quad\ell=2s_{max}-1.
\eeq
If the value of the maximal spin in the supermultiplet is set to $1/2$ we get two non-zero components of the wave function $\varphi^{-7,1}(x^{+2,0},v^{\al+,0i})$ and $\varphi^{-6,0\, a}(x^{+2,0},v^{\al+,0i})$ corresponding to the component fields of $N=2$ hypermultiplet. For $s_{max}=1$ one obtains three component fields on $S^3$ $\varphi^{-8,\,2}$, $\varphi^{-7,1\, a}$ and $\varphi^{-6,\,0\, a[2]}$ that correspond to the component fields of $N=2$ Maxwell supermultiplet. Similarly setting $s_{max}=2$ yields the on-shell field strengths of $N=2$ supergravity multiplet $(W_{\al(4)}, \Phi_{\al(3)}\vp{\Phi}^a, F_{\al(2)})(x^{m})$ \cite{Cremmer}. Choosing other values of the maximal spin gives massless higher-spin multiplets of $D=5$ $N=2$ supersymmetry. The model (\ref{superparticle-action}) is also capable to describe $D=5$ $N=4$ supergravity multiplet for $q=7/4$ and $\ell=2$. Corresponding components of the superparticle's wave function and associated fields on the space-time are listed in the Table.
\begin{center}\begin{tabular}{|c|c|}
\hline
Fields on $\vp{\hat S}S^3$ & Fields on space-time \\ \hline\hline
$\varphi^{-10,4}$ & $\vp{\wh W}W_{\al(4)}$ \\ \hline
$\varphi^{-9,\,3\:a}$ & $\vp{\wh\Phi}\Phi_{\al(3)}\vp{\Phi}^a$ \\ \hline
$\varphi^{-8,\,2\:a[2]}$ & $\vp{\tilde{\wh F}}F_{\al(2)}\vp{F}^{a[2]}$ \\ \hline
$\varphi^{-7,1}\vp{\varphi}_a$ & $\vp{\wh\Psi}\Psi_{\al a}$ \\ \hline
$\varphi^{-6,\,0}$ & $\vp{\wh S}S$ \\
\hline
\end{tabular}
\end{center}

\section{Conclusion}

Gauge invariant curvatures of YM and gravitational fields in $5d$ are known to admit equivalent spinor representation similarly to corresponding $4d$ fields. In particular, $D=5$ Riemann tensor amounts to the set of four irreducible (multi-index) spinors. We analyzed restrictions imposed by the dynamical equations and Bianchi identities on the curvature spinors for YM, gravitational and free massless spin-$3/2$ fields. In analogy with the $4d$ case in the absence of sources there remain non-zero only symmetric curvature spinors with $2s$ indices given by the Weyl curvature spinor and its lower-spin counterparts. These spinors satisfy first-order equations reducing in the linearized limit to Dirac-type equations that can be written in the uniform way for various spins. This suggests that they are the first members of the sequence of equations for higher-spin Weyl curvature spinors. These equations are solved using the integral representation based on writing the null-momentum as the square of spinors that are blocks of the spinor harmonic matrix parametrizing the coset $SO(1,4)/(SO(1,1)\times ISO(3))$ being the realization of the $S^3$ manifold and the integration is actually performed over this three-sphere. This integral representation for on-shell Weyl curvature spinors is the $D=5$ extension of that in dimensions $D=3,4,6,10$ described in \cite{DelducGS}. The possibility to elaborate such an integral representation for Weyl curvature spinors in various dimensions and for fields of various spins is due to fact that only Lorentz symmetry is manifest. So it would be interesting to look for generalizations to the fields over curved backgrounds, particularly such as $(A)dS$ ones, and in other dimensions including the string/M-theoretic ones. Less straight-forward but potentially more promising is to promote proposed superparticle model to the string one, whose correlation functions would reproduce (tree-level) scattering amplitudes in $D=5$ YM/gravity theories similarly to the (ambi)twistor-string models \cite{Witten03}, \cite{Berkovits}, \cite{Skinner}, \cite{Mason-Skinner}, \cite{Geyer}. The fact that spinorial (vectorial) constituents are Lorentz harmonics makes feasible also a generalization to other dimensions like in the case of the ambitwistor string of Ref.~\cite{Mason-Skinner}, in which Lorentz harmonics enter implicitly \cite{Bandos'14}.

\section*{Acknowledgements}

The author is obliged to A.A.~Zheltukhin for stimulating discussions.

\section*{Appendix A. $\g-$matrices and spinors}

$\g-$matrices $\g^{m}\vp{\g}_{\al}\vp{\g}^{\beta}$ ($m=0,1,2,3,5$, $\al,\bt=1,2,3,4$) satisfy the defining relations of the $D=5$ Clifford algebra
\beq
\g^{m}\vp{\g}_{\al}\vp{\g}^{\de}\g^{n}\vp{\g}_{\de}\vp{\g}^{\bt}+\g^{n}\vp{\g}_{\al}\vp{\g}^{\de}\g^{m}\vp{\g}_{\de}\vp{\g}^{\bt}=2\eta^{mn}\de_{\al}^{\bt}.
\eeq
Positions of the spinor indices can be changed with the aid of the charge conjugation matrices $C^{\al\bt}$ and $C_{\al\bt}$: $C^{\al\de}C_{\de\bt}=\de^\al_\bt$ according to the rule
\beq
\psi^\al=C^{\al\bt}\psi_\bt,\quad\chi_\al=C_{\al\bt}\chi^\bt.
\eeq
Both charge conjugation and $\g-$matrices are antisymmetric in 5 dimensions.

Subsequent relations are widely used in the main text
\beq
(C^{\al\bt})^\ast=-C_{\al\bt},\quad(\g^{m})^\dagger=\g^0\g^{m}\g^0,\quad(\g^{m}\vp{\g}_{\al}\vp{\g}^{\bt})^T=-\g^{m\bt}\vp{\g}_{\al},
\eeq
where $\g^{m\bt}\vp{\g}_{\al}=-C^{\bt\g}\g^{m}\vp{\g}_{\g}\vp{\g}^{\de}C_{\de\al}$ and minus sign reflects antisymmetry of the charge conjugation matrices.

In discussion of the Lorentz harmonics used is the light-cone basis for $\g-$matrices, in which only $SO(1,1)\times SU(2)$ covariance is manifest. Let 0 and 5 be the light-cone directions, then $\g-$matrices exhibit the direct product structure
\beq
\g^0\vp{\g}_{\mu}\vp{\g}^{\nu}=\rho^{\mr{t}}\vp{\rho}_{(\mu)}\vp{\rho}^{(\nu)}\de_i^j,\quad\g^I\vp{\g}_{\mu}\vp{\g}^{\nu}=\rho^{\mr{ts}}\vp{\rho}_{(\mu)}\vp{\rho}^{(\nu)}\tau^I\vp{\tau}_i\vp{\tau}^j,\quad\g^5=\rho^{\mr{s}}\vp{\rho}_{(\mu)}\vp{\rho}^{(\nu)}\de_i^j.
\eeq
Non-relativistic Pauli matrices satisfy
\beq
\tau^I\vp{\tau}_i\vp{\tau}^k\tau^J\vp{\tau}_k\vp{\tau}^j+\tau^J\vp{\tau}_i\vp{\tau}^k\tau^I\vp{\tau}_k\vp{\tau}^j=2\de^{IJ}\de_i^j.
\eeq
$\rho^{\mr{t}}$ and $\rho^{\mr{s}}$ are timelike and spacelike $\g-$matrices in $D=1+1$ dimensions
\beq
\rho^{\mr{t}}\vp{\rho}_{(\mu)}\vp{\rho}^{(\nu)}=\left(
\begin{array}{cc}
0 & i\\[0.2cm]
i& 0
\end{array}
\right),
\quad
\rho^{\mr{s}}\vp{\rho}_{(\mu)}\vp{\rho}^{(\nu)}=\left(
\begin{array}{cc}
0 & i\\[0.2cm]
-i& 0
\end{array}
\right)
\eeq
and $\rho^{\mr{ts}}=\rho^{\mr{t}}\rho^{\mr{s}}$. Similarly charge conjugation matrices acquire the direct product form
\beq
C^{\mu\nu}=C^{(\mu)(\nu)}\varepsilon^{ij},\quad C_{\mu\nu}=C_{(\mu)(\nu)}\varepsilon_{ij},
\eeq
where antisymmetric unit matrices $\varepsilon^{ij}$ and $\varepsilon_{ij}$:
\beq
\varepsilon^{12}=\varepsilon_{21}=1,\quad (\varepsilon^{ij})^\dagger=\varepsilon_{ji},\quad\varepsilon^{ij}\varepsilon_{jk}=\de^i_k
\eeq
are used to move $SU(2)$ indices as $\psi^i=\varepsilon^{ij}\psi_j$, $\chi_i=\varepsilon_{ij}\chi^j$, and
$D=1+1$ charge conjugation matrices equal
\beq
C^{(\mu)(\nu)}=\left(
\begin{array}{cc}
0 & i\\[0.2cm]
i& 0
\end{array}
\right),\quad C_{(\mu)(\nu)}=\left(
\begin{array}{cc}
0 & -i\\[0.2cm]
-i& 0
\end{array}
\right)
\eeq
so that
\beq
\rho^{\mr{t}}_{(\mu)(\nu)}=\left(
\begin{array}{cc}
1 & 0\\[0.2cm]
0& 1
\end{array}
\right),\quad
\rho^{\mr{s}}_{(\mu)(\nu)}=\left(
\begin{array}{cc}
1 & 0\\[0.2cm]
0& -1
\end{array}
\right)
\eeq
as desired for expressing light-cone components of the vector Lorentz harmonics in terms of the spinor ones.

\section*{Appendix B. Some properties of $D=5$ Lorentz harmonics}

Discussion of the properties of the spinor harmonics $v^{\al\boldsymbol{\mu}}$ in this section complements that in the main text.

Spinor harmonics satisfy generalized Majorana condition
\beq\label{gener-Maj-cond}
\bar v^{\alpha\boldsymbol{\mu}}=(v_{\beta\boldsymbol{\nu}})^\dagger\g^0\vp{\g}_{\beta}\vp{\g}^{\alpha}\g^0\vp{\g}_{\boldsymbol{\nu}}\vp{\g}^{\boldsymbol{\mu}}=(C^{\alpha\beta}C^{\boldsymbol{\mu\nu}}v_{\beta\boldsymbol{\nu}})^T=(v^{\alpha\boldsymbol{\mu}})^T
\eeq
reducing the number of (real) independent components from 32 to 16.
Other forms of this condition are
\beq\label{gener-Maj-cond-2}
(v_{\alpha\boldsymbol{\mu}})^\ast=\g^{0\alpha}\vp{\g}_{\beta}\g^{0\boldsymbol{\mu}}\vp{\g}_{\boldsymbol{\nu}}v^{\beta\boldsymbol{\nu}},\quad
(v^{\alpha\boldsymbol{\mu}})^\ast=\g^0\vp{\g}_{\alpha}\vp{\g}^{\beta}\g^0\vp{\g}_{\boldsymbol{\mu}}\vp{\g}^{\boldsymbol{\nu}}v_{\beta\boldsymbol{\nu}}
\eeq
can be used to check reality of the vector harmonics realized in terms of the spinor ones (\ref{vector-vs-spinor-harmonics}).

The light-cone basis realization for $D=5$ $\g-$matrices implies decomposition of $Spin(1,4)$ spinor on $SO(1,1)\times SU(2)$ representations $\mathbf4=\mathbf2_+\oplus\mathbf2_-$. Applied to the index that transforms under $SO(1,4)_R$ introduces decomposition of the spinor harmonic matrix on $4\times2$ blocks (cf. Eq.~(\ref{spinor-harmonics}))
\beq\label{component-spinor-harmonics}
v^{\alpha\boldsymbol{\mu}}=v^{\alpha(\mu)i}=\left(
\begin{array}{c}
v^{\alpha+i}\\
v^{\alpha-i}
\end{array}
\right),\quad v^{\alpha}\vp{v}_{(\mu)i}=C_{(\mu)(\nu)}\varepsilon_{ij}v^{\alpha(\nu)j}=-i\left(
\begin{array}{c}
v^{\alpha-}\vp{v}_i\\
v^{\alpha+}\vp{v}_i
\end{array}
\right).
\eeq
For such blocks generalized Majorana condition reduces to the $SU(2)-$Majorana condition
\beq
(v^{\alpha\,\pm i})^\ast=\g^0\vp{\g}_{\alpha}\vp{\g}^{\beta}v_{\bt}\vp{v}^{\pm}\vp{v}_i.
\eeq

\end{document}